# On the Implementation of a Reinforcement Learning-based Capacity Sharing Algorithm in O-RAN

I. Vilà, O. Sallent, J. Pérez-Romero
Dept. of Signal Theory and Communications, Universitat Politècnica de Catalunya (UPC)
Barcelona, Spain
irene.vila.munoz@upc.edu, sallent@tsc.upc.edu, jordi.perez-romero@upc.edu

*Abstract*— **The capacity sharing problem in Radio Access Network (RAN) slicing deals with the distribution of the capacity available in each RAN node among various RAN slices to satisfy their traffic demands and efficiently use the radio resources. While several capacity sharing algorithmic solutions have been proposed in the literature, their practical implementation still remains as a gap. In this paper, the implementation of a Reinforcement Learning-based capacity sharing algorithm over the O-RAN architecture is discussed, providing insights into the operation of the involved interfaces and the containerization of the solution. Moreover, the description of the testbed implemented to validate the solution is included and some performance and validation results are presented.**

## I. Introduction

The vision of 5G as a system that simultaneously supports multiple new services (e.g., virtual reality, smart cities, etc.) with heterogeneous requirements (e.g., high bit rate, low latency, reliability) motivated the specification of network slicing as a key feature of the 5G architecture. This feature allows sharing a common network infrastructure among multiple communication providers, referred to as tenants, by providing each one of them with an end-to-end logical network, i.e., network slice, optimized to its specific requirements.

The deployment of network slices in the Radio Access Network (RAN), i.e., RAN slices, involves the provisioning of multiple and diverse RAN behaviours over the common and scarce pool of radio resources at each RAN node. To achieve this, mechanisms that allow dynamically distributing the capacity available at each RAN node are fundamental. The challenge is to perform the capacity sharing according to the traffic demands of the different RAN slices while satisfying the requirements established in the Service Level Agreement (SLA) for each RAN slice and, at the same time, achieving an efficient use of the available capacity.

Several works can be found in the literature that have proposed capacity sharing solutions, ranging from heuristic and optimization-based solutions, such as [1] and [2], to Reinforcement Learning (RL)-based solutions [3] and [4]. Whereas the existing works have focused on the formulation and assessment of algorithmic solutions, to the best of authors' knowledge none of them has focused on their practical implementation. In this paper, an implementation framework for capacity sharing solutions within the O-RAN architecture is firstly introduced. Based on this, the paper presents an O-RAN compliant implementation of a specific RL-based capacity sharing algorithm, namely, the Deep Q-Network - Multi-Agent Reinforcement Learning (DQN-MARL) capacity sharing solution from our previous work [4] by describing the required interfaces and involved protocols for the interaction between the DQN-MARL capacity sharing solution and the RAN nodes as well as the containerization of the solution.

The implementation framework of this paper is embraced within the scope of the PORTRAIT project [5], whose main objective is to conduct the proof of concept (PoC) of the DQN-MARL solution in a real environment in the context of the 5GCAT pilot [6], which is part of the 5G Spanish National Plan. The 5GCAT pilot includes an O-RAN compliant field trial 5G small cell network deployment in a beach close to Barcelona city. As a preliminary stage for the PoC, the testbed presented in this paper has been developed in the laboratory to validate the implemented algorithm as well as the necessary interfaces that should allow a nearly plug-and-play integration into the 5GCAT pilot platform.

The rest of the paper is organized as follows. Section II introduces the key elements of the O-RAN architecture that are relevant for the implementation of capacity sharing solutions. Section III describes the implementation of the DQN-MARL capacity sharing solution. Then, Section IV includes the description of the testbed developed to validate the proposed implementation and Section V provides the validation and performance results obtained with the testbed. Finally, Section VI summarizes the conclusions.

## II. Capacity Sharing within O-RAN Architecture

O-RAN Alliance was launched in 2018 with the aim to standardize a RAN architecture that complements the 3GPP architecture with a set of open interfaces for the realization of a virtualized RAN with disaggregated functionalities and embedded Artificial Intelligence (AI) [7]. Given that O-RAN architecture is expected to be adopted in numerous 5G deployments supporting multiple slices, the implementation of capacity sharing solutions on top of this architecture deserves a proper analysis for assessing their practical feasibility.

The O-RAN logical architecture is depicted in Fig. 1, where the key components for the implementation of capacity sharing solutions are highlighted in blue. The architecture is composed of disaggregated O-RAN functions and open interfaces as well as 3GPP interfaces [8]. The Service Management and Orchestration (SMO) function is responsible for the management of the rest of O-RAN functions and the O-Cloud. These are the O-RAN Central Unit – Control Plane (O-CU-CP), the O-RAN Central Unit – User Plane (O-CU-UP), the O-RAN Distributed Unit (O-DU), the O-RAN Radio Unit (O-RU) and the O-RAN eNB (O-eNB). Four key interfaces are introduced to enable interoperation between the SMO and the rest of O-RAN functions and the O-Cloud. These are the A1, O1, Open Fronthaul (FH) M-plane and O2 interfaces.

To perform closed-loop optimization control and orchestrate the RAN with enhanced AI-powered functionalities, two RAN Intelligent Controllers (RIC) have been included in the architecture. First, the near-real-time RIC (near-RT RIC) is deployed in the edge of the network operating in control loops with a periodicity between 10 ms and 1 s by interacting with the O-CU-CP/UP and O-DU via the E2 interface. Second, the non-real-time RIC (non-RT RIC) is deployed at the SMO. It enables a closed-loop control of the RAN and SMO operations with time scales larger than 1 s. The non-RT RIC supports the execution of third-party applications, referred to as *rApps*, to provide value-added services that support and facilitate policy management, RAN analytics and machine learning model management and to deliver enriched information through the A1, O1 and O2 interfaces. *rApps* can support functionalities such as frequency and interference management, capacity sharing or SLA assurance [7]. The support to these functionalities is provided through the O1 interface that allows configuring the parameters in the O-RAN functions according to gathered performance measurements of the status of the network.

The capacity sharing functionality can be implemented in the O-RAN architecture as an *rApp* that decides on the allocated capacity to each slice in each cell, where a cell provides coverage in a certain area on a given frequency carrier and is associated with a O-CU, O-DU and O-RU. This can be performed by configuring the parameter *rRMPolicyDedicatedRatio* of the O-DU function for each cell. This parameter belongs to the 3GPP Network Resource Model (NRM) for characterizing RRM Policies [9] and specifies the percentage of radio resources that can be dedicatedly used by a slice, which is identified by the Single Network Slice Assistance Information (S-NSSAI). The *rApp* for capacity sharing makes decisions on the value of *rRMPolicyDedicatedRatio* based on gathered performance measurements from the O-DU function. Note that this use case has been identified by the O-RAN use case specification for the non-RT RIC in [10].

## III. IMPLEMENTATION OF THE CAPACITY SHARING *RAPP*

This section describes the implementation of a specific RL-based capacity sharing solution, namely, the DQN-MARL algorithm for multi-tenant and multi-cell scenarios presented in [4]. The implemented *rApp* corresponds to the inference stage of the DQN-MARL capacity sharing solution, i.e., the application of the actions in the real network according to previously learnt policies. The *rApp* is implemented as a container that includes all the elements for the interaction with the O-DUs via the O1 interface. This is detailed in the following.

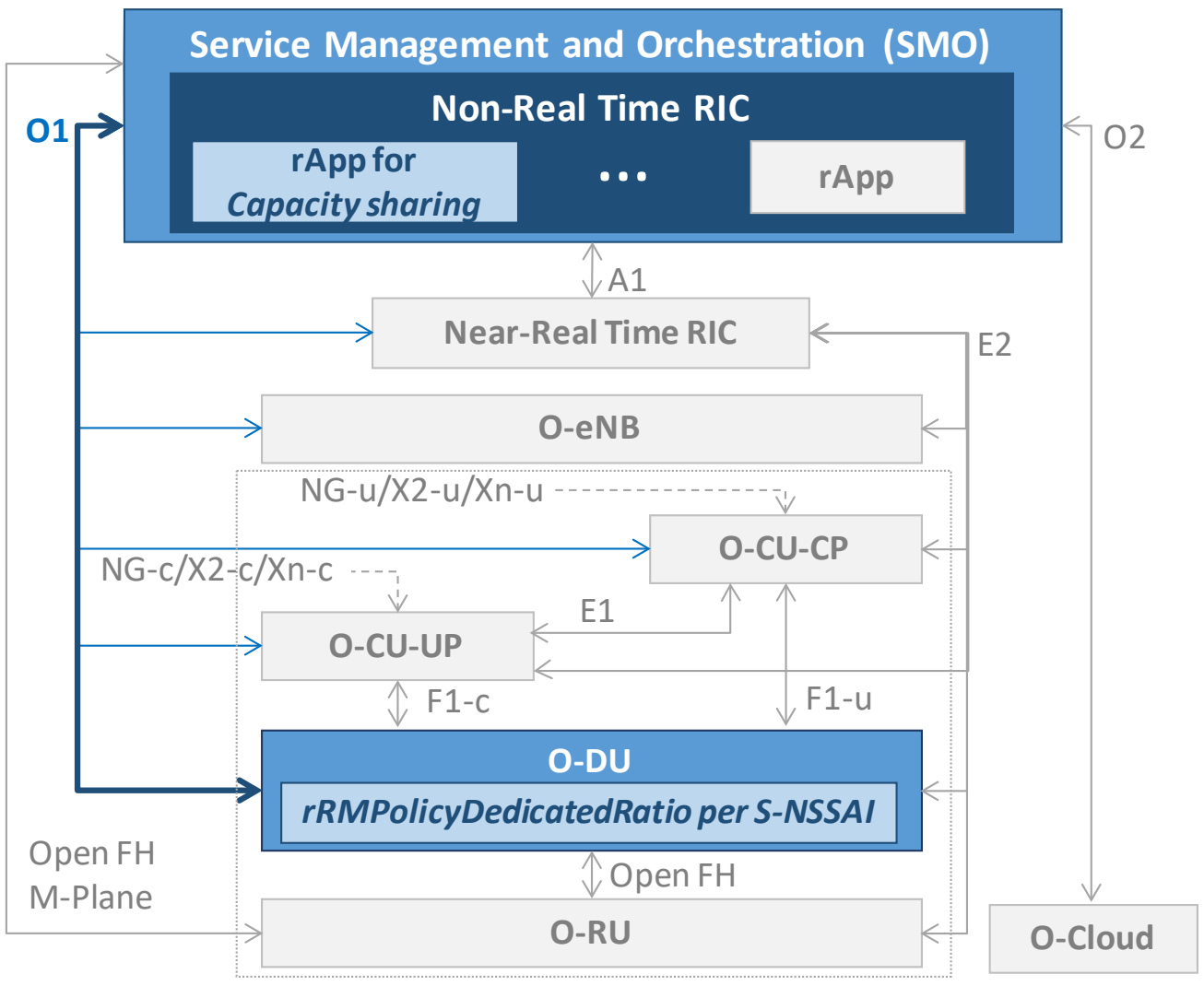

Fig. 1. O-RAN architecture with capacity sharing implementation.

### A. *O1 interface implementation*

The O1 interface enables the interworking between the *rApp* and each one of the O-DUs that handle the cells controlled by the capacity sharing *rApp*, as depicted in Fig. 2. The *rApp* acts as Management Service (MnS) Consumer of the O1 interface while the O-DU acts as MnS producer. Two MnSs need to be deployed: the *Provisioning MnS* to configure the *rRMPolicyDedicatedRatio* attribute per S-NSSAI at the O-DU, and the *Performance Assurance MnS* to gather the performance measurements from the O-DU [11].

The *Provisioning MnS* relies on the NETCONF protocol. NETCONF defines a simple mechanism through which a network device (e.g., an O-DU) can be managed, configuration data can be retrieved, and new configuration data can be uploaded and manipulated [12]. It has a layered architecture, where the core is a Remote Procedure Call (RPC) layer transported over secure transports such as Secure Shell (SSH), Transport Layer Security (TLS), Simple Object Access Protocol (SOAP) or Blocks Extensible Exchange Protocol (BEEP), being SSH mandatory. The protocol operates according to a service-client scheme, where the MnS Producer has the role of the server and the MnS Consumer the role of the client. Then, the implemented capacity sharing *rApp* includes a NETCONF client for each O-DU handling a cell in order to establish the NETCONF connection with the NETCONF server at the O-DU.

Moreover, NETCONF defines a set of base protocol operations. One of them is the edit-config operation that allows the MnS Consumer to modify the configuration of parameters at the MnS Producer through a XML-encoded file, following the procedure in Fig. 3. Once the NETCONF session is established, the MnS Consumer sends an RPC request to edit the configuration of the Managed Object Instance (MOI) attributes at the MnS Producer. Attributes are defined by using the YANG data modelling language that allows expressing the structure and semantics of configuration information in a vendor-neutral format and in a readable and compact way. After sending the RPC request, the MnS Producer replies whether confirming the MOI configuration or notifying an error, and finally, the NETCONF connection terminates. Using this procedure, the *rApp* for capacity sharing (MnS Consumer) can send modification requests including the desired value of the *rRMPolicyDedicatedRatio* attribute per S-NSSAI to the O-DU function (MnS Producer) through XML files. To support this configuration, the NETCONF server includes the standardised 3GPP YANG modules that define the *rRMPolicyDedicatedRatio* per

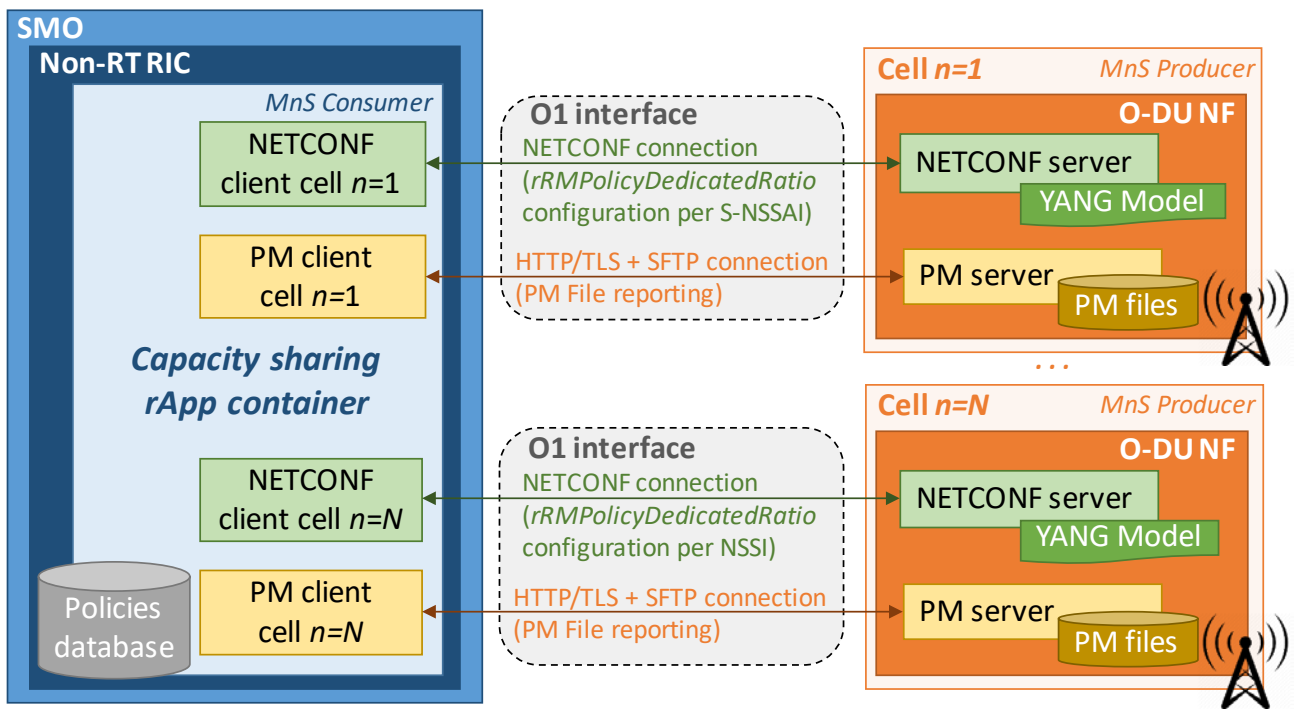

Fig. 2. DQN-MARL Capacity sharing implementation.

S-NSSAI as an attribute of an object named RRMPolicyRatio. These modules are available in [9].

The *Performance Assurance MnS* allows the O-DU (MnS Producer) to report Performance Measurements (PM) data to the capacity sharing *rApp* (MnS consumer). This can be done through either the PM data file mode (PMs are obtained with periodicities in the order of minutes) or the PM Data streaming mode (PMs are obtained in real time). The implemented solution considers the PM data file mode since capacity sharing operates in the order of minutes. Accordingly, PM files are obtained by following the procedure in Fig. 4. Whenever a new PM data file is available, the MnS Producer sends an asynchronous *notifyFileReadyNotification* to the MnS Consumer over HTTP/TL with the name and location of the file. Then, the MnS Consumer can retrieve the file from the specified location through SFTP or STPeS. The PM data file is an XML-based file defined according to the 3GPP formats of [13] and the PM definitions of [14].

To support the *Performance Assurance MnS* and establish the HTTP/TL and SFTP connections, the implemented capacity sharing solution contains a PM client for each O-DU, which retrieves PM files from the PM server at the O-DU. The PM server collects measurements and generates PM files with the PM metrics required by the DQN-MARL capacity sharing solution. Specifically, following the 3GPP specifications in [15], the collected PM data per cell that are needed by the DQN-MARL solution are the "Mean DL PRB used for data traffic", which measures the average number of Physical Resource Blocks (PRB) used for data traffic per S-NSSAI, the "DL total available PRB", which provides the average number of available PRBs in the downlink, and the "DL PDCP PDU Data Volume", which provides the downlink data volume delivered from the O-CU to the O-DU per S-NSSAI and is used for the throughput computation in the solution.

### B. *Containerization*

The implementation of the *rApp* for capacity sharing uses containerization, so that it can run as an isolated software unit called container that packages up all the required code, libraries, binaries, configuration files and dependencies required by the *rApp* to run [16]. In this way, the *rApp* container can be encapsulated and easily deployed on the SMO/non-RT RIC platform, where the *rApp* runs isolated from the rest of containers while all of them share the operating system (OS) and are managed by a container engine such as Docker.

*Algorithm* 1 summarizes the operation of the code that has been developed for the operation of the capacity sharing *rApp* container that implements the inference stage of the DQN-MARL solution. After establishing the NETCONF and HTTP/TLS sessions between the O-DU of each cell, the policies of each tenant (i.e., S-NSSAI) are loaded from the policy database (lines 1-3 of *Algorithm* 1). Then, the *notifyFileReady* notification is received periodically from the PM servers at the O-DUs and the PM files are transferred to the *rApp* container via SFTP. The data in the PM file is used to compute the state of each tenant (lines 5-7 of *Algorithm* 1)., Then, the *rRMPolicyDedicatedRatio* per S-NSSAI is computed for each cell by applying the corresponding policy to the computed state. The resulting *rRMPolicyDedicatedRatio* is sent as a NETCONF modify attribute request (lines 8-9 of *Algorithm* 1). The operations in lines 5-9 of *Algorithm* 1 are repeated every $\Delta t$ minutes. The code of *Algorithm* 1 has been developed in Python using the library TF-agents. This code is included in the capacity sharing *rApp* container jointly with the required dependencies for the operation of this code, the access to the policy database, the required version of Python and the libraries to run the code.

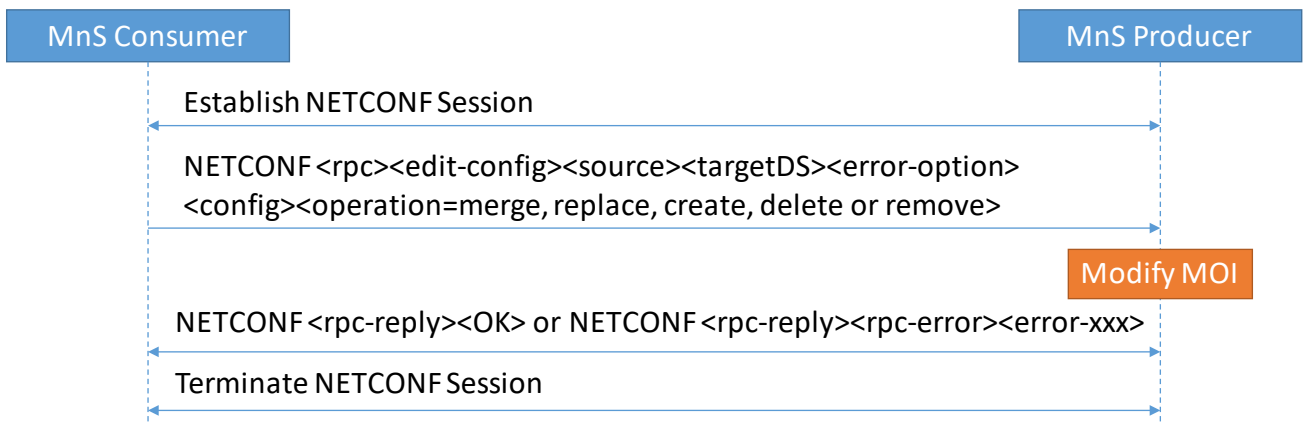

Fig. 3. Simplified Modify MOI Attributes procedure.

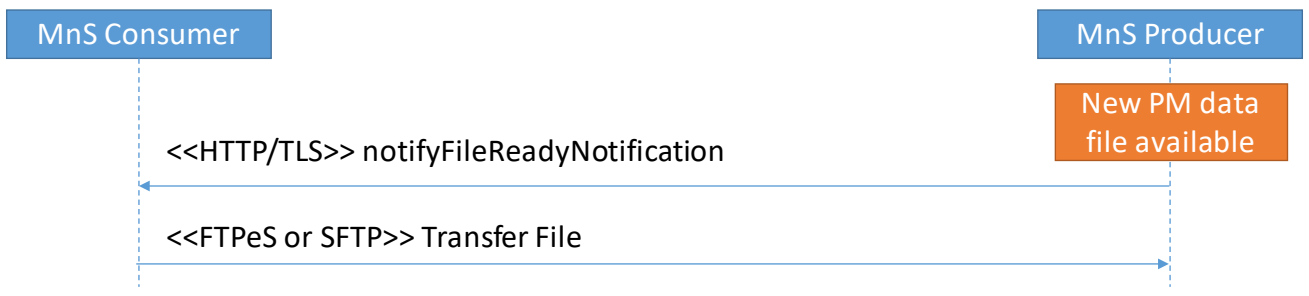

Fig. 4. PM Data File Reporting.

**_Algorithm_ 1 – Capacity sharing _rApp_ container operation**

1 Establish NETCONF session for each O-DU
2 Establish HTTP/TLS session for each O-DU
3 Load saved policy from policies database.
4 Periodically (Loop):
5 Receive *notifyFileReady* via HTTP/TLS from each cell.
6 Obtain PM measurements through SFTP for each cell.
7 Compute state from PM files for each S-NSSAI.
8 Obtain the *rRMPolicyDedicatedRatio* for each S-NSSAI and each cell by applying the policies.
9 Send *rRMPolicyDedicatedRatio* per S-NSSAI via NETCONF modify attribute request for each cell.

## IV. DEVELOPED TESTBED

The testbed depicted in Fig. 5 has been developed and set up at the laboratory to validate the operation of the *rApp* and the O1 interface. The testbed, which considers a scenario of a single cell, is composed of two hosts denoted as Host 1 and Host 2 in a private Local Area Network (LAN) with IP addresses 192.768.0.1 and 192.768.0.2, respectively. Host 1 represents the SMO/Non-RT RIC side in Fig. 2 and contains an *rApp* for capacity sharing running as a Docker container. The *rApp* has been developed in Python according to *Algorithm* 1 and uses previously trained policies as inputs. Moreover, an SLA monitoring container is also included in Host 1. It stores performance data and provides data visualization of the SLA assurance to the Tenant through a Graphical User Interface (GUI).

Host 2 represents the O-DU in Fig. 2 and contains the required elements of the O1 interface at the O-DU side. In this regard, a NETCONF server and a PM SFTP Server are included, both running as Docker containers. They allow testing the O1 interface through the interaction with the NETCONF and PM SFTP clients in the *rApp* in Host 1. The NETCONF server container has been built using the available NETCONF server container image at [17] that is part of the ONAP NF simulator. This image allows loading YANG modules and performing operations for reading configurations (get-config operation) and editing them via XML files (edit-config operation). In order to configure the *rRMPolicyDedicatedRatio* per S-NSSAI, the NETCONF server has been provided with the standardised YANG module for the RRMPolicyRatio object called " 3gpp-nr-nrm-rrmpolicy.yang", which is available in [9] and includes the *rRMPolicyDedicatedRatio* as an attribute. Moreover, the O-DU contains a SFTP server to store PM measurement files and make them available.

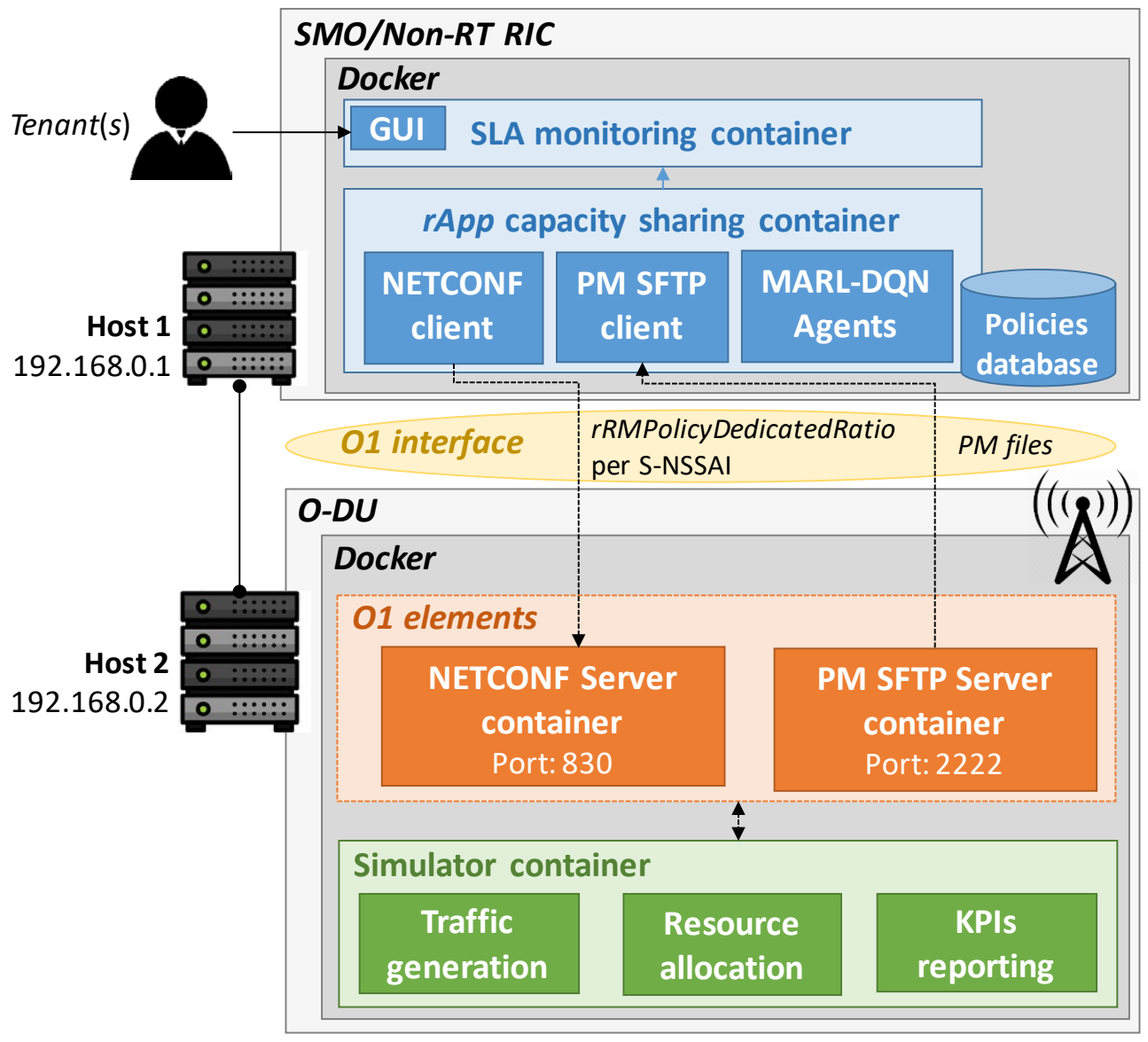

Fig. 5. Testbed design

Host 2 also includes a Docker container with a simulator for generating the traffic of different tenants and for performing the resource allocation to each tenant during the simulation time according to the values of *rRMPolicyDedicatedRatio* per S-NSSAI stored in the NETCONF server container. Based on this, the simulator generates PM files every $\Delta t$ minutes and stores them at the PM SFTP Server. These PM files are defined according to the standardized formats in the XML schema file "measData.xsd" available in the 3GPP specification in [13] and include all the required measurements by the *rApp* at Host 1 to compute the state and perform the inference of the trained policies. For the reader's information, an open software repository has been created that includes the software solution of the DQN-MARL algorithm, as well as the elements required for implementation of the O1 interface included in the testbed. The software repository is available in the GitHub platform at [18]. In addition to the source code of the different elements of the testbed, the repository provides a series of tutorial-like documentation that guides the user through the set up and execution of the different elements of the implemented solution.

## V. Validation and Performance Results

This section presents first the considered scenario for obtaining results using the developed testbed. Then, validation results of the O1 interface interaction of the *rApp* in the testbed and the evaluation of the *rApp* performance are provided.

### A. Considered scenario

The *rApp* for capacity sharing has been validated on the testbed in Fig. 5 by considering a relevant 5G scenario with two tenants, denoted as Tenant 1 and Tenant 2. Tenant 1 provides enhanced Mobile BroadBand (eMBB) services and Tenant 2 provides a Fixed Wireless Access (FWA) service to home users. FWA provides a wireless alternative to wired broadband connection and is one of the first 5G use cases that has generated increased momentum [19].

Both tenants are served by a single cell with capacity $c_n$= 117 Mb/s, as it is deployed in the testbed in Fig. 5. The SLA established for the $k$-th tenant is defined in terms of the Scenario Aggregated Guaranteed Bit Rate, $SAGBR_k$, to be provided across all cells to the tenant if requested and the Maximum Cell Bit Rate, $MCBR_k$, that can be provided to the tenant in each cell. Specifically, the established SLAs for Tenant 1 and 2 are $SAGBR_1$=70.2 Mb/s and $SAGBR_2$=46.8 Mb/s, corresponding to the 60% and the 40% of the cell capacity $c_n$, respectively, and $MCBR_1$= $MCBR_2$=93.6 Mb/s, corresponding to the 80% of $c_n$. These scenario parameters are set in the simulator container and configured in the capacity sharing *rApp* container.

To obtain the policies to be applied in the testbed the DQN-MARL capacity sharing solution in [4] has been trained. To this end, the solution considers a different DQN agent per tenant that learns the policy that tunes the *rRMPolicyDedicatedRatio* so that the SLA of the tenant is satisfied and an efficient use of the capacity in the cell is achieved. The tuning of the *rRMPolicyDedicatedRatio* is performed in time steps of duration $\Delta t$=3 min. At each time step, the DQN agent of the $k$-th tenant obtains the state from the network environment and, accordingly selects an action, which can be to increase/decrease the value *rRMPolicyDedicatedRatio* of the cell in an action step of $\Delta$=3% or to keep it equal. As a result, a reward is obtained assessing the quality of the selected action. By performing this interaction iteratively, the policy of each tenant is updated until a converged policy is obtained. For further details on the state, action and reward definitions and the training process, the reader is referred to [4].

The training has been performed using a training dataset composed of several offered load patterns of the two tenants until reaching convergence with the hyperparameters of the DQN-MARL model shown in Table I. Once the training has finished, the learnt policies have been stored in the policies database in Fig. 5 to be used by the *rApp* capacity sharing container during the inference stage in the testbed.

### B. O1 interface validation

This section presents the validation of the interactions between the capacity sharing *rApp* container and the O-DU through the O1 interface developed in the testbed. Fig. 6 depicts the sequence of the most relevant messages during the execution of a single time step (i.e., a single loop in *Algorithm* 1) obtained from the logging of the *rApp* at Host 1. The figure shows the time stamps of the different messages in the format (seconds, milliseconds).

The interaction between the PM SFTP client in the *rApp* at Host 1 and the PM SFTP server at Host 2 to obtain PM files occurs from time 32,487 to time 32,961. The first messages until time 32,636 show the procedure to establish the SFTP connection, which consists in firstly establishing an SSH connection, executing the exchange of keys, the user authorisation, and the establishment of a *sesch channel* for the SFTP client (*rApp* Host 1) to execute commands in the SFTP server (PM SFTP server in Host 2). After this process, the SFTP connection is established, and the *rApp* can search the PM files in the SFTP server (from time 32.926 to 32.932) and download them (at time 32.946). After this, the SFTP session is closed. Note that these messages included correspond to the "*SFTP Transfer File* " in Fig. 4.

The PM file contains the PM measurements of both tenants in the considered scenario. The *rApp* processes the downloaded PM file to compute the state, which is provided to the DQN-MARL algorithm to obtain the *rRMPolicyDedicatedRatio* per S-NSSAI, as shown in Fig. 6 from times 32,961 until 33,040.

TABLE I. DQN-MARL MODEL HYPERPARAMETERS FOR TRAINING

| Parameter | Value |
|---|---|
| Initial collect steps | 50 |
| Maximum number of time steps for training | $2 \cdot 10^5$ |
| Experience Replay buffer maximum length ($l$) | $10^7$ |
| Mini-batch size ($J$) | 512 |
| Learning rate ($\tau$) | 0.0001 |
| Discount factor($\gamma$) | 0.9 |
| $\varepsilon$ value ($\varepsilon$-Greedy) | 0.1 |
| DNN configuration | Input layer: 7 nodes<br>1 full connected layer: 100 nodes<br>Output layer: 9 nodes |

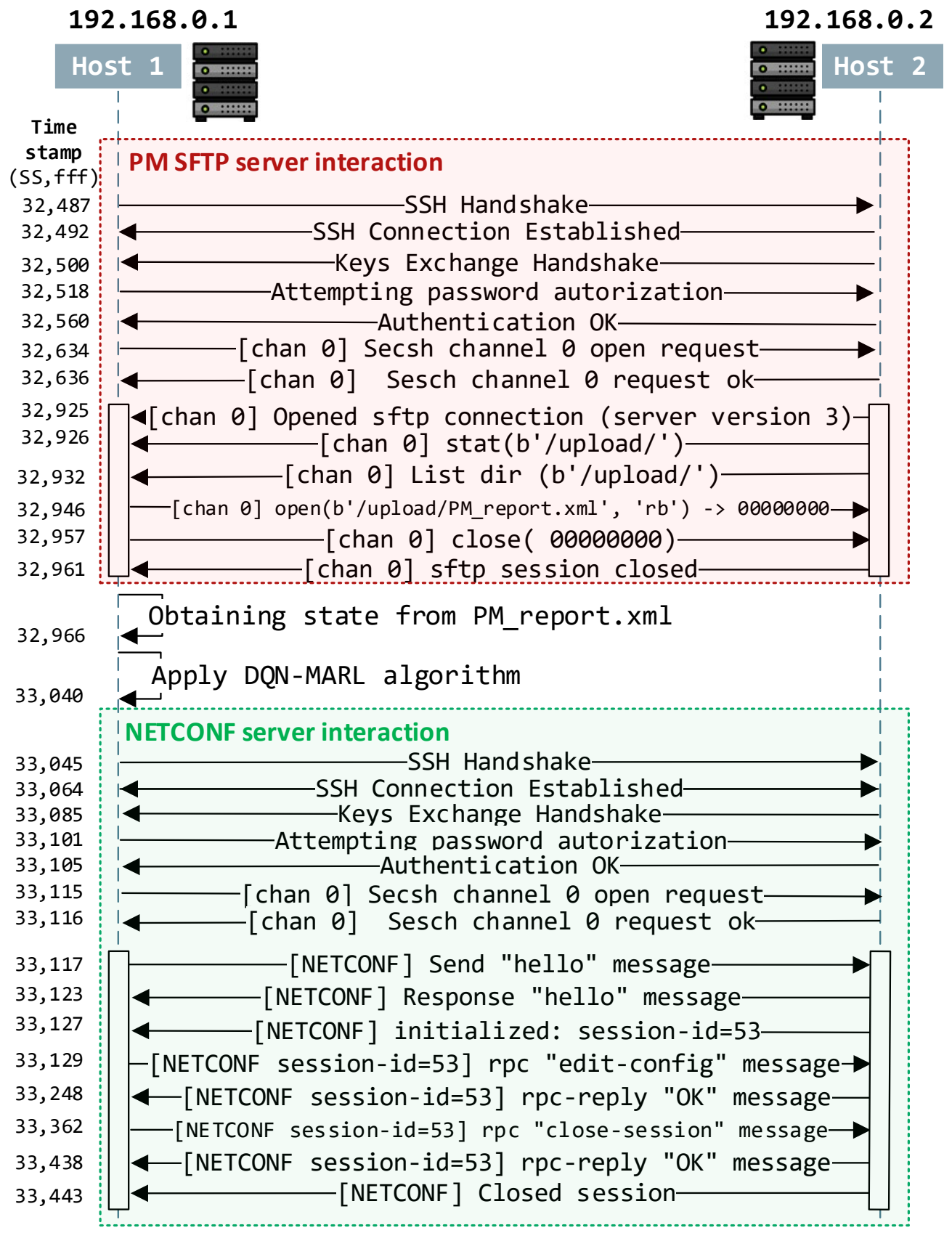

Fig. 6. O1 interface interaction.

After this, the NETCONF client in the capacity sharing *rApp* container at Host 1 and the NETCONF server at Host 2 interact in order to update the *rRMPolicyDedicatedRatio* per S-NSSAI at the NETCONF server with the new obtained values. The messages corresponding to this interaction are those from time 33,045 until the end in Fig. 6 and they implement the procedure specified in Fig. 3. Firstly, as the NETCONF connection considered here builds upon a secure SSH connection, this connection is established from time 33,045 to 33,116. Then, the *rApp* and the NETCONF server at Host 2 perform a "hello" messages handshake (at times 33.117 and 33.123), providing their NETCONF capabilities (i.e. configurations and operation modes) that determine the configuration of the established NETCONF session (received at time 33,127). After this, the *rApp* sends the new *rRMPolicyDedicatedRatio* configurations through the RPC "edit-config" message at time 33,129. The detail of this message formatted in XML is included in Fig. 7, which sets the *rRMPolicyDedicatedRatio* of Tenant 1 (i.e., id=1) to 57% and the *rRMPolicyDedicatedRatio* of Tenant 2 (i.e. id=2) to 42%. As a result of this "edit-config" message, the NETCONF

```xml
<?xml version="1.0" encoding="UTF-8"?>
<nc:rpc message-id="urn:uuid:edb2c826-1bb6-4837-b3f3-57524d8d31d3"
xmlns:nc="urn:ietf:params:xml:ns:netconf:base:1.0">
 - <nc:edit-config>
   - <nc:target>
       <nc:running/>
     </nc:target>
     <nc:test-option>set</nc:test-option>
   - <nc:config>
     - <RRMPolicyRatio xmlns="urn:3gpp:sa5:_3gpp-nr-nrm-rrmpolicy">
         <id>1</id>
       - <attributes>
           <rRMPolicyDedicatedRatio>57</rRMPolicyDedicatedRatio>
         </attributes>
       </RRMPolicyRatio>
     - <RRMPolicyRatio xmlns="urn:3gpp:sa5:_3gpp-nr-nrm-rrmpolicy">
         <id>2</id>
       - <attributes>
           <rRMPolicyDedicatedRatio>42</rRMPolicyDedicatedRatio>
         </attributes>
       </RRMPolicyRatio>
     </nc:config>
   </nc:edit-config>
</nc:rpc>
```

Fig. 7. RPC "edit-config" message in XML format.

server sends an rpc-reply "OK" message at time 33.248 confirming that the *rRMPolicyDedicatedRatio* values have been successfully updated and finally, the NETCONF session closes with the RPC "close-session" handshake from time 33,438 until 33,443.

The provided sequence of messages proves that the *rApp* is able to successfully interact through the O1 interface to obtain PM files and send NETCONF configurations. The reader is referred to the Github platform in [18] for further details on the O1 interaction, where the source code is available together with a sample PM file obtained from the SFTP server.

## C. Performance evaluation

In this section, the performance of the *rApp* for capacity sharing is assessed on the testbed in Fig. 5 when the traffic generated in the simulator container by the two tenants corresponds to the one depicted in Fig. 8. The figure shows the evolution of the offered load (i.e., the fraction of the cell capacity required by the tenant) of Tenant 1, $o_1(t)$, and Tenant 2, $o_2(t)$, during a week as well as the total offered load in the cell, $o(t)$. The offered load of Tenant 1, providing an eMBB service, experiences large values during the morning from Monday until Friday, while lower offered load values are obtained during the weekend. In turn, the offered load of Tenant 2 exhibits a typical behaviour of a FWA service, increasing during the last hours of the day when people arrive at home. Note that the resulting total offered load $o(t)$ during working days, exceeds the cell capacity sometimes during the afternoon.

The *rApp* for capacity sharing has been tested for the offered loads in Fig. 8 according to the procedure in *Algorithm* 1. As a result, the SLA monitoring container provides the result of Fig. 9, which compares the evolution of offered load of both tenants with the evolution of the configured *rRMPolicyDedicatedRatio* for Tenant 1 and Tenant 2, denoted as $rRMPolicyDedicatedRatio_k(t)$ for $k$=1,2. Results show how $rRMPolicyDedicatedRatio_k(t)$ ratios for both tenants generally adapt to the offered loads during all the week, providing more than the $SAGBR_k$ if required as long as the total offered load does not exceed the cell capacity. For example, this is observed in the peaks of Monday to Friday for Tenant 1, in which the value of $rRMPolicyDedicatedRatio_1$ can be higher than $SAGBR_1$ =60% because the offered load of Tenant 2 is much lower than $SAGBR_2$ =40%. However, when the total offered load exceeds the system capacity, $rRMPolicyDedicatedRatio_k(t)$ ratios lower than the offered loads are experienced. For instance, this occurs for the offered load of Tenant 1 at middays during the working days or for Tenant 2 at night when FWA demands are high. However, when this situation occurs, at least the $SAGBR_k$ of the tenant is provided, guaranteeing the SLA.

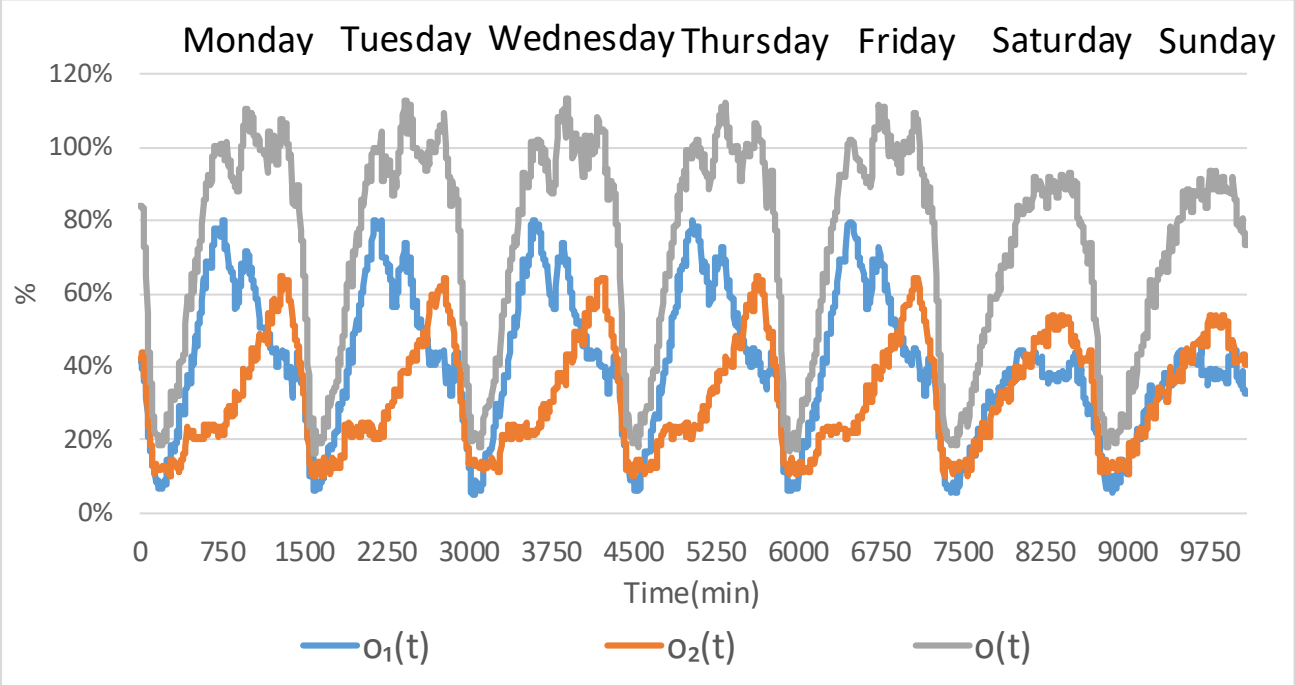

Fig. 8. Offered load evolution of both tenants during a week.

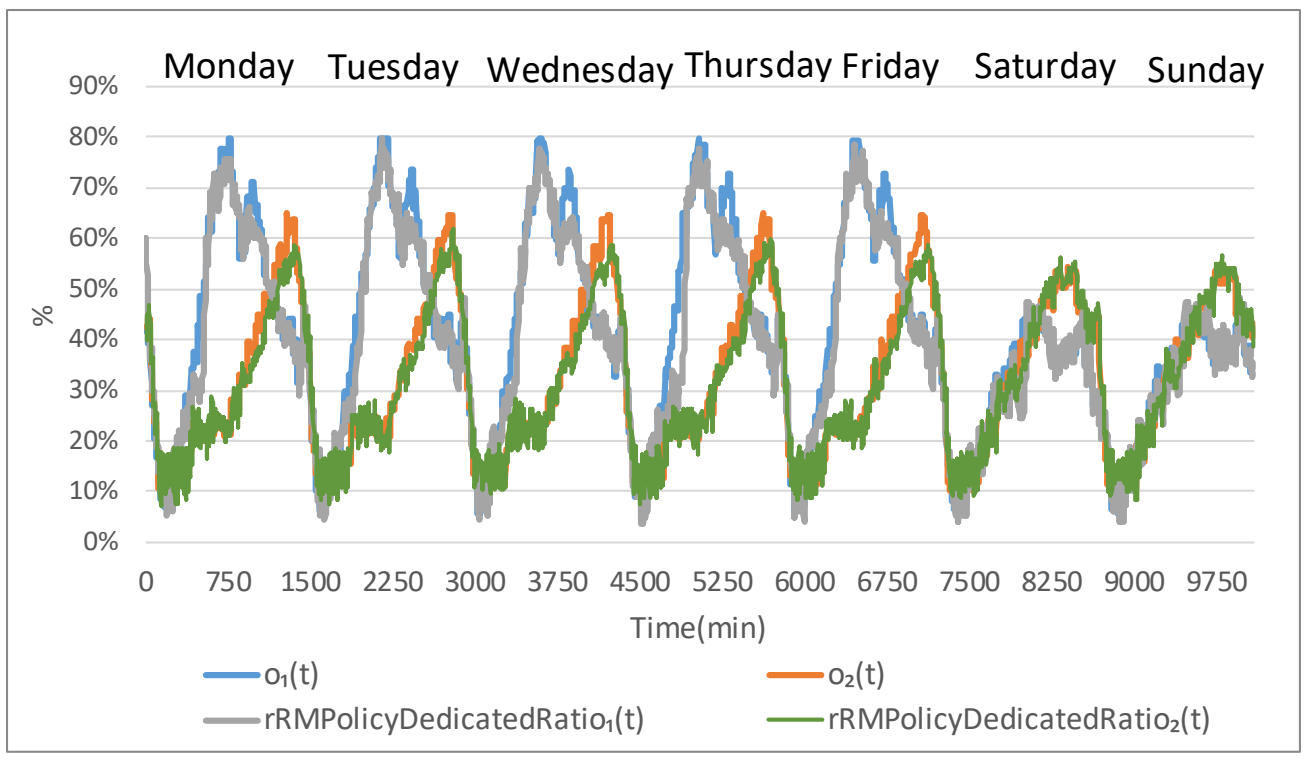

Fig. 9. Offered load vs $rRMPolicyDedicatedRatio_k(t)$ evolution.

The SLA monitoring container also provides the average satisfaction ratio per tenant, computed as the average number of time steps that the SLA is satisfied with respect to the total number of time steps. The obtained values during the period of Fig. 9 for Tenant 1 and Tenant 2 are 0.94 and 0.96, respectively, reflecting a high degree of SLA fulfilment. The average assigned capacity utilization is also computed by the SLA monitoring container. This metric measures the amount of capacity overprovisioning and is computed as the average of the ratio between the used capacity by both tenants and the assigned capacity to both tenants. The resulting capacity utilization is 0.84, which shows that the assigned capacity is generally not overprovisioned. These results highlight the capability of the solution to efficiently distribute the capacity in the cell among the two tenants providing eMBB and FWA services according to their offered loads while satisfying the SLAs and without capacity overprovisioning.

## VI. Conclusions and Future Work

This paper has described the implementation of a RL-based capacity sharing algorithm for RAN slicing. To ensure the O-RAN compliance of the solution, the algorithm has been developed as an *rApp* and the O1 interface required for the interaction of the solution with the RAN nodes of the 5G network has been implemented.

The solution has been containerized and a testbed has been built in the laboratory. This testbed has allowed validating the interaction of the *rApp* through the O-RAN interfaces and assessing the behaviour of the solution in a relevant 5G use case. Performance results have shown that the *rApp* is able to adapt to the traffic demands and achieve high service level agreement satisfaction ratios above 0.94 with a minimum overprovisioning. Moreover, an open source software repository that includes the different elements of the solution and their corresponding documentation has been provided in Github as an illustrative practical example for the scientific community in this field.

As future work, it is planned to integrate the developed *rApp* in a real environment in the context of the 5GCAT pilot that includes a field trial 5G small cell network deployment.

## Acknowledgment

This paper is part of PORTRAIT project (ref. PDC2021-120797-I00) funded by MCIN/AEI/10.13039/501100011033 and by European Union Next GenerationEU/PRTR.

## References

[1] A. S. D. Alfoudi, et. al "An Efficient Resource Management Mechanism for Network Slicing in a LTE Network," in *IEEE Access*, vol. 7, pp. 89441-89457, 2019.

[2] J. Pérez-Romero, et.al , "Profit-Based Radio Access Network Slicing for Multi-tenant 5G Networks," 2019 *Europ. Conf. on Networks and Comms. (EuCNC)*, Valencia, Spain, 2019, pp. 603-608.

[3] R. Li et al., "Deep Reinforcement Learning for Resource Management in Network Slicing," in *IEEE Access*, vol. 6, pp. 74429-74441, 2018.

[4] I. Vilà, J. Pérez-Romero, O. Sallent, A. Umbert, "A Multi-Agent Reinforcement Learning Approach for Capacity Sharing in Multi-tenant Scenarios," in *IEEE Trans. Veh. Tech.*, vol. 70 no. 9, July 2021.

[5] "PORTRAIT (Proof Of concept of a Radio access neTwoRk slicing solution based on Artificial InTelligence) project". Accessed 12 July 2022 from https://portrait.upc.edu/.

[6] 5GCAT. "The decisive drive for a digital society". Accessed 11 July 2022 from https://pilot5gcat.com/en/.

[7] M. Polese, et. al. "Understanding O-RAN: Architecture, Interfaces, Algorithms, Security, and Research Challenges," arXiv:2202.01032 [cs.NI], 2022. [Online].

[8] O-RAN.WG1.O-RAN-Architecture-Description-v06.00, "O-RAN Architecture Description version 6.00," O-RAN Alliance, Working Group 1, Technical specification, Nov. 2021.

[9] 3GPP TS 28.541 v16.0.0, "Management and orchestration; 5G Network Resource Model (NRM) (Release 16)," March 2019.

[10] O-RAN.WG2.Use-Case-Requirements-v05.00, "O-RAN Non-RT RIC & A1 Interface: Use Cases and Requirements version 5.00", Working Group 2, Technical Specification, Nov. 2021.

[11] O-RAN.WG10.O1-Interface.0-v05.00, "O-RAN Operations and Maintenance Interface Specification," O-RAN Alliance, Working Group 10, Technical Specification, August 2020.

[12] J. Schönwälder, et. al. "Network configuration management using NETCONF and YANG," in *IEEE Communications Magazine*, vol. 48, no. 9, pp. 166-173, Sept. 2010.

[13] 3GPP TS 28.550 v16.7.0, "Management and orchestration; Performance assurance (Release 16)," Dec. 2020.

[14] 3GPP TS 28.554 v16.5.0, "Management and Orchestration; 5G end to end Key Performance Indicators (KPI) (Release 16)", July, 2020.

[15] 3GPP TS 28.552 v16.2.0, "Management and orchestration; 5G Performance measurements (Release 16)," June 2019.

[16] A. Gopalasingham, et. al. "Virtualization of radio access network by Virtual Machine and Docker: Practice and performance analysis," 2017 *IFIP/IEEE Symp. on Int. Net. and Serv. Manag. (IM)*, 2017.

[17] ONAP, "NF Simulator". Accessed 27 January 2022 from https://docs.onap.org/projects/onap-integration/en/latest/simulators/nf_simulator.html.

[18] I.Vilà, J. Pérez-Romero, O. Sallent, A. Umbert, "DQN-MARL Capacity Sharing Solution", Github. Available at: https://github.com/grcmupc/portrait.

[19] Ericsson, "Fixed Wireless Access handbook", extracted version, 4rth edition, 2021. Available at: ericsson.com/fwa.